\documentclass[twocolumn]{pasj01}
\jyear{2024}
\Received{$\langle$reception date$\rangle$}
\Accepted{$\langle$acception date$\rangle$}
\Published{$\langle$publication date$\rangle$}
\usepackage{url}
\usepackage[T1]{fontenc}
\usepackage{ae,aecompl,rotating}
\usepackage[switch,mathlines]{lineno}
\usepackage{graphicx}
\usepackage{color}

\begin{document}

\title{Combining neural networks with galaxy light subtraction for discovering strong lenses in the HSC SSP}

\author{Yuichiro \textsc{Ishida}\altaffilmark{1,2}}
\email{yuichiro.ishida@grad.nao.ac.jp}

\author{Kenneth C. \textsc{Wong}\altaffilmark{3,4}}

\author{Anton T. \textsc{Jaelani}\altaffilmark{5,6}}

\author{Anupreeta \textsc{More}\altaffilmark{7,8}}

\altaffiltext{1}{Department of Astronomy, The University of Tokyo, 7-3-1 Hongo, Bunkyo-ku, Tokyo 113-0033, Japan}
\altaffiltext{2}{Department of Earth and Planetary Sciences, Kyushu University, 744 Motooka, Nishi-ku, Fukuoka 819-0395, Japan}
\altaffiltext{3}{Research Center for the Early Universe, Graduate School of Science, The University of Tokyo, 7-3-1 Hongo, Bunkyo-ku, Tokyo 113-0033, Japan}
\altaffiltext{4}{National Astronomical Observatory of Japan, 2-21-1 Osawa, Mitaka, Tokyo 181-8588, Japan}
\altaffiltext{5}{Astronomy Research Group and Bosscha Observatory, FMIPA, Institut Teknologi Bandung, Jl. Ganesha 10, Bandung 40132, Indonesia}
\altaffiltext{6}{U-CoE AI-VLB, Institut Teknologi Bandung, Jl. Ganesha 10, Bandung 40132, Indonesia}
\altaffiltext{7}{The Inter-University Centre for Astronomy and Astrophysics (IUCAA), Post Bag 4, Ganeshkhind, Pune 411007, India}
\altaffiltext{8}{Kavli IPMU (WPI), UTIAS, The University of Tokyo, Kashiwa, Chiba 277-8583, Japan}


\maketitle
\begin{abstract}
Galaxy-scale strong gravitational lenses are valuable objects for a variety of astrophysical and cosmological applications. Strong lensing galaxies are rare, so efficient search methods, such as convolutional neural networks, are often used on large imaging datasets.  In this work, we apply a new technique to improve the performance of supervised neural networks by subtracting the central (lensing) galaxy light from both the training and test datasets. We use multiband imaging data from the Hyper Suprime-Cam Subaru Strategic Program (HSC SSP) as our training and test datasets.  By subtracting the lensing galaxy light, we increase the contrast of the lensed source compared to the original imaging data.  We also apply the light subtraction to non-lenses in order to compare them to the light-subtracted lenses. Residual features resulting from poor light subtraction can adversely affect the performance of networks trained on the subtracted images alone.  We find that combining the light-subtracted images with the original $gri$-band images for training and classification can overcome this and improve the overall classification accuracy.  We find the area under the receiver operating characteristic curve can be improved to 0.841 using the combination of the fiducial images and light-subtracted images, compared to 0.808 for the fiducial imaging dataset alone.  This may be a promising technique for improving future lens searches using CNNs.
\end{abstract}


\section{Introduction} \label{sec:intro}
Galaxy-scale strong gravitational lenses are valuable objects for a variety of astrophysical and cosmological applications, including studies of the mass structure of galaxies (e.g., \cite{koopmans+2006,sonnenfeld+2013,shajib+2021}), the properties of dark matter (e.g., \cite{more+2009,vegetti+2014,hezaveh+2016,nierenberg+2017,gilman+2020}), resolved studies of distant sources at high resolution (e.g., \cite{rybak+2015,canameras+2017}), and the measurement of cosmological parameters (e.g., \cite{refsdal1964,wong+2020,birrer+2020}).

However, strong lensing galaxies are rare, requiring a chance alignment of a foreground galaxy (the ``lens", often a massive elliptical galaxy) with a bright background object (the ``source", often a star-forming galaxy or a quasar).  As a result, the best datasets for discovering strong lenses are deep, wide-field, multiband imaging surveys that cover a large area of the sky.  For example, the Sloan Digital Sky Survey (SDSS; e.g., \cite{bolton+2006}), the Kilo-Degree Survey (KiDS; e.g., \cite{li+2021, he+2020, petrillo+2017}) and the Hyper Suprime-Cam Subaru Strategic Program (HSC SSP; e.g., \cite{sonnenfeld+2018,sonnenfeld+2020,chan+2020,chan+2024,wong+2022}) provide such datasets, sometimes with supplementary spectroscopic data that can aid lens searches.  The Vera C. Rubin Observatory's Legacy Survey of Space and Time (LSST; \cite{ivezi+2019}) will eventually cover a majority of the sky to unprecedented depths.  Therefore, efficient methods are needed to search for strong lenses in these large datasets, such as arc-finding algorithms (e.g., \cite{more+2012}), citizen science searches (e.g., \cite{more+2016,sonnenfeld+2020}), and machine learning methods such as convolutional neural networks (CNNs).

CNNs can process imaging data efficiently and identify characteristic patterns.  CNNs have been proven to be able to classify strong lensing objects with reasonable accuracy in multiband imaging data (e.g., \cite{jacobs+2017,jacobs+2019,petrillo+2019,huang+2020,li+2020}), substantially reducing the number of lens candidates that need to be visually inspected to verify them.  The main challenge in identifying galaxy-scale gravitational lenses is distinguishing the multiple lensed images from the foreground galaxy light, as the image separation tends to be comparable to the atmospheric seeing, particularly for lenses with small Einstein radius.  Using multiband data is helpful, as the sources tend to be blue star-forming galaxies while the lenses tend to be red elliptical galaxies, but faint sources can still be challenging for CNNs to identify.

One way to increase the contrast between the lens galaxy and the lensed images is to subtract the lens galaxy light from the images.  The RINGFINDER algorithm \citep{gavazzi+2014} used lens light subtraction, which helped to detect many small angular separation lenses that were missed by citizen scientists who were not able to view such subtracted images (see Figure 7 of \cite{more+2016}).  A subsequent citizen science search by \citet{sonnenfeld+2020} used the {\sc YattaLens} software \citep{sonnenfeld+2018} to subtract the central galaxy light, which helped to improve the discovery of these lenses, demonstrating the effectiveness of light subtraction for this purpose.

In this work, we attempt to improve the performance of lens-finding CNNs by subtracting the central galaxy light from both the training and test datasets of our networks using {\sc YattaLens} to highlight the light from the lensed source compared to the original imaging data.  A similar technique was tried by \citet{canameras+2023}, who used difference images between multiband observations as a proxy for lens light subtraction.  We use a more sophisticated method to subtract the central galaxy light, but compare our results qualitatively to those of \citet{canameras+2023}.  Such light subtraction has generally not been applied to CNN-based lens searches in the past, but may be important for upcoming ground-based surveys (e.g., LSST).  We also apply the subtraction to non-lenses in order to make the conditions the same, as we expect an ideal light subtraction to leave no residual features since there is no background source.  Therefore, our neural networks can more easily find the characteristics of the source galaxy and classify strong lensing objects correctly.  We use multiband imaging from the Wide layer of the HSC SSP as our training and test data.  The HSC SSP is an ideal dataset for discovering strong lenses due to its depth and area, and many galaxy-scale strong lenses have already been discovered as part of the Survey of Gravitationally lensed Objects in HSC Imaging (SuGOHI\footnote{https://www-utap.phys.s.u-tokyo.ac.jp/~oguri/sugohi/}; \cite{sonnenfeld+2018,wong+2018,wong+2022,chan+2024}) project.  Several studies, including some working with the public HSC SSP data, have made use of neural networks to discover lens candidates in this dataset \citep{canameras+2021,shu+2022,andika+2023,jaelani+2023,schuldt+2024}.

We construct three datasets consisting of the same objects for training: the fiducial HSC $gri$-band imaging dataset, the dataset with light subtraction applied, and the combination of the two. We first optimize our CNN for the fiducial dataset, train it on the other two datasets, and compare the performance of each model.  Then, we build new models optimized for each dataset and compare their performance in order to determine whether the light subtraction is able to improve the ability of CNNs to identify strong lenses.

This paper is organized as follows.  We describe the data used in this study and our galaxy light subtraction procedure in Section~\ref{sec:data}.  We describe our neural network design and architecture in Section~\ref{sec:cnn}.  We present the performance of our networks when applying our light subtraction procedure and evaluate the characteristics of correctly and incorrectly classified objects in Section~\ref{sec:results}.  We summarize our conclusions and discuss potential future work in Section~\ref{sec:summary}.  Throughout this paper, all magnitudes given are on the AB system.

\section{Data} \label{sec:data}

\subsection{HSC SSP imaging} \label{subsec:hsc_data}
The HSC SSP is a wide-field imaging survey conducted with the HSC instrument \citep{miyazaki+2012,miyazaki+2018,kawanomoto+2018,komiyama+2018} on the Subaru telescope.
The data used in this paper are from the Wide component of the HSC SSP, which consists of $grizy$ imaging (although we only use the $gri$ bands) to a depth of $z\sim26.2$ at a pixel scale of $0\farcs168$/pixel.  The imaging data used in this paper are taken from Public Data Release 2 (PDR2; \cite{aihara+2019}) of the HSC SSP, although we use updated variance images and point spread functions (PSFs) from Data Release 4 (DR4).

\subsection{Training and test data} \label{subsec:training_data}
The training data for this study consist of mock lenses and real non-lenses taken from the training sample of \citet{jaelani+2023}, which we briefly summarize here.

The mock lenses were generated using {\sc simct}\footnote{https://github.com/anumore/SIMCT} \citep{more+2016}, which superimposes simulated lensed arcs on top of image cutouts of massive galaxies from HSC.  The galaxy mass models are assumed to be singular isothermal ellipsoids, and the velocity dispersion $\sigma$ is obtained from the $L - \sigma$ relation of \citet{parker+2005}.  The ellipticity and position angle of the mass is assumed to follow that of the light distribution.  External shear drawn from a uniform distribution is added to the mass model.  The source galaxies are parameterized as a S\'ersic profile with an index of 1.  Their redshift and magnitude distributions are drawn from \citet{faure+2009}, and their ellipticity and position angle is drawn from a uniform distribution.  The source sizes are estimated from the $L-r_{\mathrm{eff}}$ relation of \citet{bernardi+2003} with an additional term to account for redshift evolution, and the colors are taken from the CFHTLenS catalog \citep{hildebrandt+2012,erben+2013}.  The lensed arcs are then generated using {\sc gravlens} \citep{keeton2001}.

The non-lenses are all real HSC objects and consist of a random sample of galaxies from the PDR2 parent catalog, along with spiral galaxies from \citet{tadaki+2020}.  A small number of stars ($\sim5\%$ of all the non-lenses) and visually-inspected ``tricky" galaxies ($\sim2\%$) are added to the sample.  These systems consist of galaxy groups, mergers, point-like galaxies, and other unusual features.

The training data used here contain 18645 mock lenses and 18642 non-lenses.  We note that the number of objects is slightly lower than the 18660 each from \citet{jaelani+2023} because a few objects did not have updated PSFs or variance images from DR4.  We use a 80/20 train/validation split for the mock lenses and the non-lenses.

We create three training datasets, which we denote {\sc fid}, {\sc sub}, and {\sc stack}, as follows:
\begin{enumerate}
\item {\sc fid}: The fiducial HSC $gri$-band imaging data (3 frames)
\item {\sc sub}: The HSC imaging data with light subtraction (3 frames)
\item {\sc stack}: The combination of {\sc fid} and {\sc sub} (6 frames)
\end{enumerate}
The selected objects in each dataset are identical, and the cutouts are 64 pixels $\times$ 64 pixels (roughly $10\farcs8\times10\farcs8$).  Our data augmentation procedure, described in Section~\ref{subsec:data_aug}, is applied independently to each of the galaxies in each dataset.

The test data, which we use to evaluate the performance of our networks, are constructed from a combination of 180 known galaxy-scale lenses and 2996 non-lenses described in \citet{more+2024}.  Specifically, we use their ``L1" and ``L2" samples, which consist of lenses discovered in HSC or KiDS using various machine learning searches, and their ``N1" sample, which consists of objects from HSC that are classified as having a low lensing probability by citizen scientists \citep{sonnenfeld+2020} in Space Warps \citep{marshall+2016,more+2016} and have been cross-checked against known SuGOHI lens candidates.

\subsection{Galaxy light subtraction} \label{subsec:lightsub}
To create the {\sc sub} dataset, we use a modified version of the {\sc YattaLens} software \citep{sonnenfeld+2018} to subtract the central galaxy light from the mock lens and nonlens samples.  Here, we briefly describe how we use the {\sc YattaLens} algorithm to subtract the galaxy light.

\begin{figure*}
    \centering
	\includegraphics[width=0.8\textwidth]{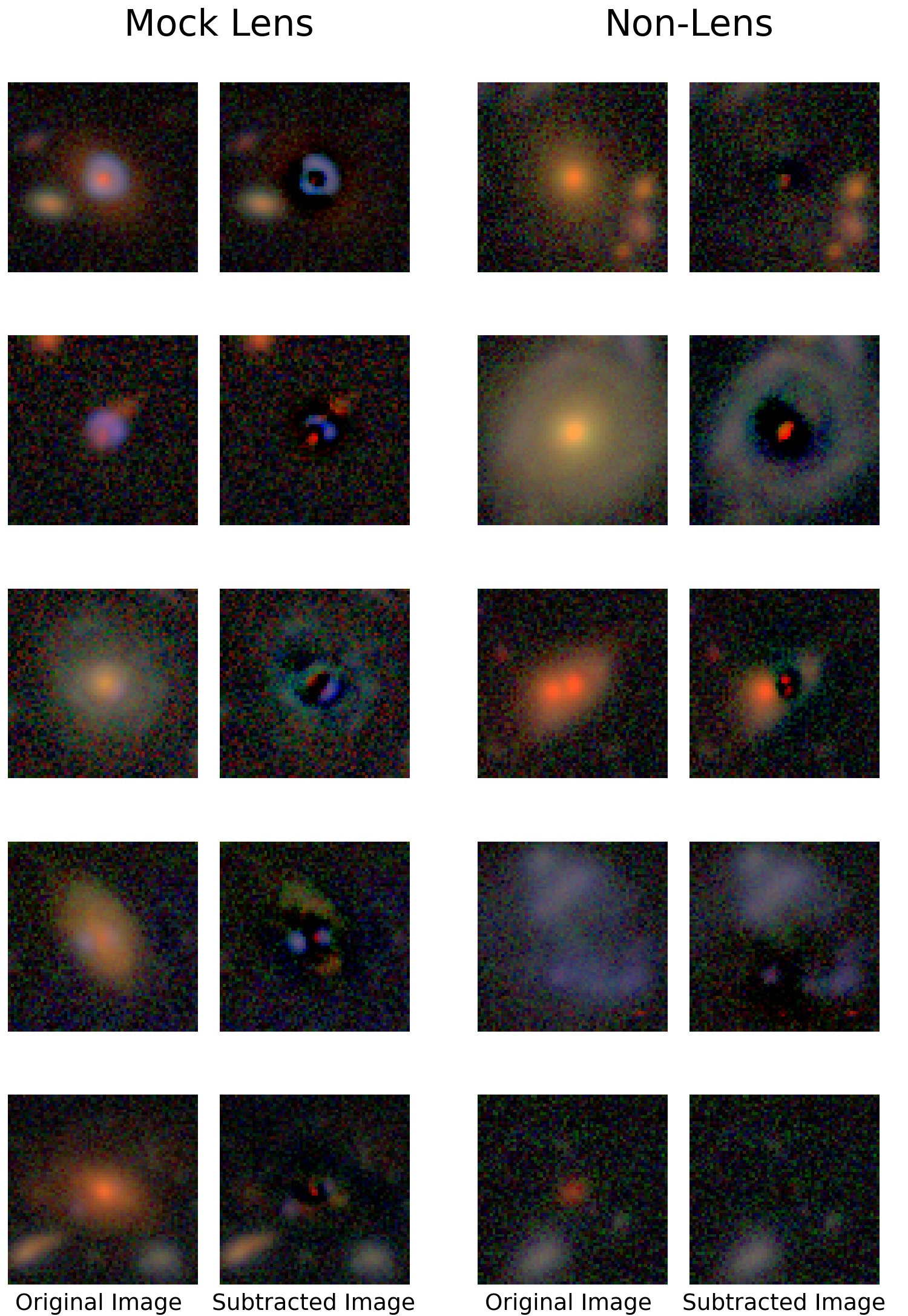}
    \caption{Comparison of the subtracted $gri$-band images of training data with the fiducial HSC images.  The objects are chosen to reflect a range of morphologies and configurations.   The left section is composed of the images of the mock lenses.  The right section is composed of the images of the non-lenses.  In each section, the left images are the fiducial images, and the right images are the same objects with the central galaxy light subtracted.  The images are 64 pixels $\times$ 64 pixels ($10\farcs8\times10\farcs8$).}
    \label{fig:compare}
\end{figure*}

\begin{figure*}
    \centering
	\includegraphics[width=0.8\textwidth]{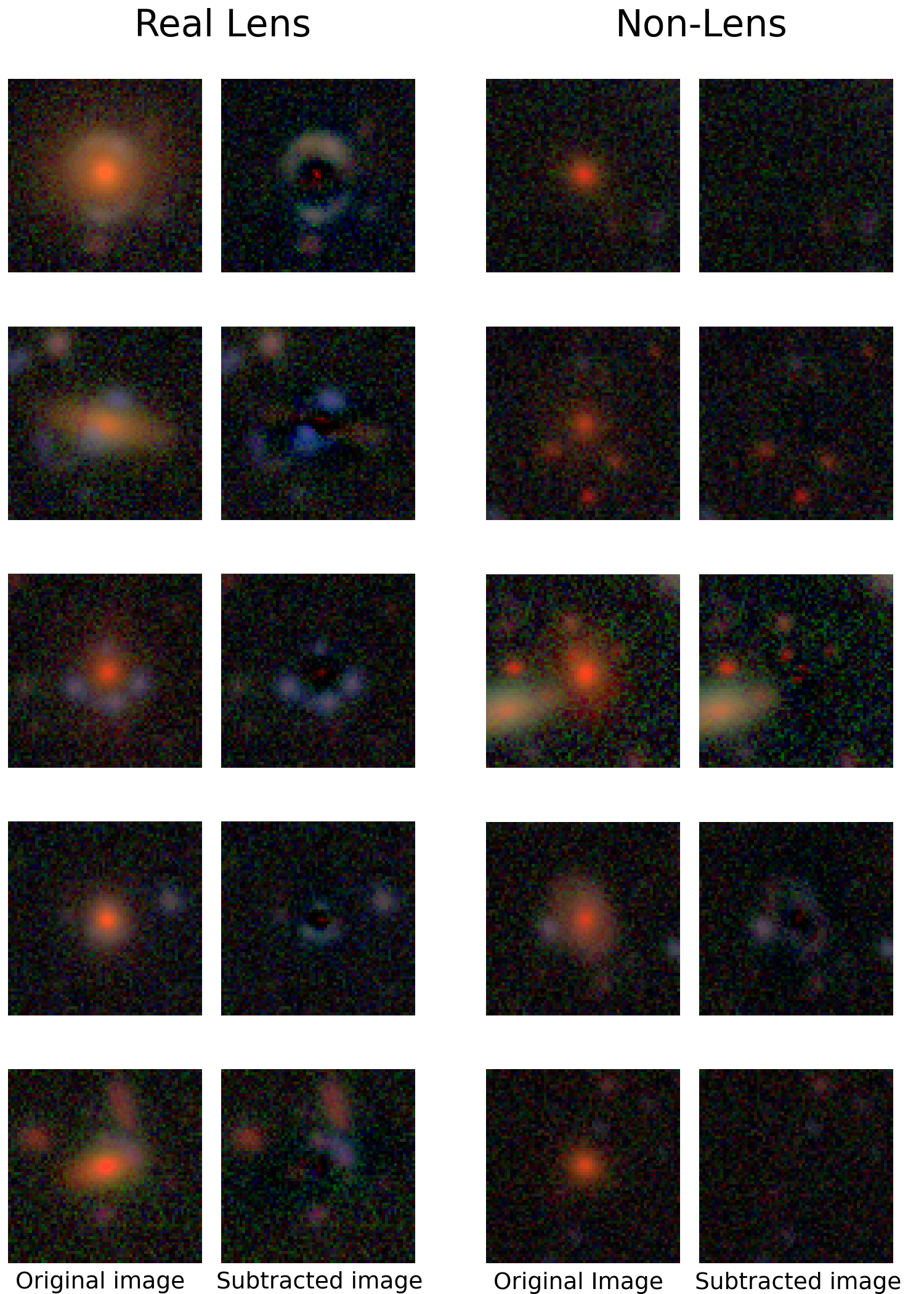}
    \caption{Comparison of the fiducial and subtracted images of the test data from HSC.  The left section is composed of the real lens images from the ``L1" and ``L2" samples of \citet{more+2024}.  The right section is composed of the non-lens images from the ``N1" sample of \citet{more+2024}. In each section, the left images are the fiducial images, and the right images are the same objects with the central galaxy light subtracted.  The images are 64 pixels $\times$ 64 pixels ($10\farcs8\times10\farcs8$).}
    \label{fig:compare_td}
\end{figure*}

{\sc YattaLens} takes as input the $gri$-band cutout images of an object, the associated variance images, and the associated PSFs.  Although the variance images for the mock lenses should account for the added Poisson noise from the added lensed images, in practice this has little effect on the central galaxy light subtraction (A. Sonnenfeld, private communication), so we use the pipeline variance images from the original cutouts.  {\sc YattaLens} fits an elliptical S\'{e}rsic profile to the $i$-band image within a 3\arcsec~radius region around the center of the galaxy.  The best-fit profile is then convolved with the PSF and subtracted from all three bands.  {\sc YattaLens} then calls {\sc SExtractor} \citep{bertinarnouts1996} to search the central galaxy-subtracted $g$-band cutout for objects that could be lensed arcs using the following criteria:
\begin{enumerate}
\item Distance from the lens center between 3~pixels and 30~pixels (roughly $0\farcs5 < R < 5\farcs0$).
\item Major-to-minor axis ratio $>$ 1.4.
\item Maximum difference of $30^{\circ}$ between the position angle of the major axis of the object and the direction tangential to a circle centered on the lens and passing through the object centroid.
\item Minimum angular aperture (the angle subtended by the object as measured from the lens centroid) of $25^{\circ}$.
\item A footprint size between 20~pixels and 500~pixels.
\end{enumerate}
Objects that satisfy all of these criteria are considered to be potential lensed arcs.  If no candidate arcs are detected, the residual images from the initial fit are used as the light-subtracted images.  If at least one candidate arc is detected, the objects detected in the $g$-band are masked out in the corresponding $i$-band image and the main galaxy is fit again with a S\'{e}rsic profile, and the residuals from this fit are used as the light-subtracted images.  Figure~\ref{fig:compare} shows a sample of fiducial images and the corresponding light-subtracted images for mock lenses and non-lenses from the training data.  Figure~\ref{fig:compare_td} shows a similar sample of objects from the test data.

Running YattaLens on the entire training sample takes roughly $\sim1000$ CPU-hours, given the specific settings we used.  We note that we made no attempt to optimize the runtime of this procedure, and the YattaLens algorithm is not necessarily the most efficient for this purpose, but this step only needs to be done once to generate the galaxy-subtracted images needed to construct the {\sc sub} and {\sc stack} datasets.  Future optimization/parallelization of YattaLens or other light subtraction algorithms may be necessary to further improve the runtime of this step for larger datasets (e.g., LSST), but the computational resource cost for this work is small in comparison to those used for training the networks.

\subsection{Image scaling and data augmentation}
\label{subsec:data_aug}
  We scale the image data using an algorithm based on \citet{lupton+2004}.  We first scale the $g$, $r$, and $i$ band images by multiplicative factors
  determined through trial and optimization and were found to give the best results (see Section~\ref{subsec:color_opt}).  We then apply the normalization described by the equations
\begin{equation}
I = \frac{g + r + i}{3} \nonumber ,
\end{equation}
\begin{equation}
(g, r, i)_\mathrm{norm} = \frac{\sinh^{-1}(e^{10}\times I)}{\sinh^{-1}(e^{10})} \times \frac{(g,r,i)}{I} + 0.05.
\end{equation}
$g$, $r$, and $i$ are the fluxes of each pixel in the respective bands, while $g_\mathrm{norm}$, $r_\mathrm{norm}$, and $i_\mathrm{norm}$ are the corresponding fluxes of each pixel after scaling.  We choose this normalization as opposed to the square-root stretch as it performs slightly better in our tests.

We use data augmentation to increase the diversity of the training data and teaching the network translational and rotational invariance.  We apply data augmentation in-place (i.e., the size of the training sample remains the same) at the start of training as follows:
\begin{enumerate}
\item a random shift ranging between -6~pixels and +6~pixels in both the x- and y-directions
\item a random adjustment of the image contrast in the range $c \in [0.9, 1.1]$. In this procedure, we adjust the flux, $f$, of each pixel to $(f - \langle f \rangle) \times c + \langle f \rangle$ for each channel, where $\langle f \rangle$ is the mean flux of the image.
\item a random scaling of the image brightness in the range [0.9, 1.1]
\item a random horizontal and vertical flip, each with 50\% probability
\item a random rotation in the range [-36,36]~deg
\end{enumerate}
The range of values chosen for the above augmentation steps are reasonable based on some initial tests we performed using early versions of our network.  Points outside the boundary of the input data are filled with zeros, as we tested other fill modes and found that they gave similar results.

\section{Neural network architecture} \label{sec:cnn}
In this section, we describe the construction of our CNN that will be trained on the various datasets.  To create an optimal network architecture, we implement the Keras Tuner software \citep{omalley+2019}, which adjusts several hyperparameters and compares different models to optimize performance.  We use the Hyperband tuner \citep{li+2016}, which compares different models by calculating the area under the receiver operating characteristic (see Section~\ref{sec:results}) from the validation data.

Our network starts with two convolutional layers each with $f_{c1}$ filters, where $f_{c1} \in [32, 64, 128]$.  We then add three groups of convolutional layers, each followed by a max pooling layer.  The convolutional layer groups consist of $2l_{c1}$,~$2l_{c2}$, and $2l_{c3}$ layers, respectively, where $l_{c1} \in [1,2,3]$ and  $l_{c2},~l_{c3} \in [1,2,3,4]$.  The convolutional layers in each of the three groups have $f_{c1}$, $f_{c2}$, and $f_{c3}$ filters, respectively, where $f_{c2},~f_{c3} \in [32, 64, 128]$ with the additional condition that $f_{c1} \leq f_{c2} \leq f_{c3}$.  We then add two dense layers with $f_{d1},~f_{d2} \in [32, 64, 128]$ filters, requiring that $f_{d2} \leq f_{d1}$.  Two boolean hyperparameters, $d_1$ and $d_2$, determine whether or not each dense layer is followed by a dropout layer with an 0.4 dropout rate.  In total, the tuner optimizes 11 hyperparameters.

The Adam optimization algorithm is chosen to minimize the binary cross-entropy error function over the training data with a learning rate of 0.001, which we decided on based on our experience with earlier versions of our network.  We use a batch size of 64 images.  Early stopping is used to save the best model to minimize the influence of overfitting if the validation loss does not improve within ten epochs.  An earlier version of our network (see Appendix) has shown to be competitive with and perhaps more flexible than those of previous studies that applied machine learning techniques to HSC SSP data \citep{holloway+2024,more+2024}.

We first train the network on the {\sc fid} dataset using the Keras Tuner to optimize the network architecture.  Then we fix the network architecture and train it on the {\sc sub} and {\sc stack} datasets to compare the results.  We subsequently try to improve the performance of the networks trained on the {\sc sub} and {\sc stack} datasets by optimizing the color stretch values and modifying the network structure using the Keras Tuner to re-optimize the hyperparameters.

We use either a Nvidia RTX A4000 (6144 CUDA cores) or Nvidia L4 (7424 CUDA cores) as the GPU accelerator for training.  It takes approximately 15 minutes to train a network with the Nvidia RTX A4000 and 5 minutes to train networks with the Nvidia L4 (not including hyperparameter optimization).
We note that training on the {\sc stack} dataset takes roughly $\sim10\%$ longer per epoch than training on the {\sc fid} or {\sc sub} datasets.  This is an inevitable consequence of having additional frames per object, but is a reasonable tradeoff if the classification accuracy is improved.

\section{Results and discussion} \label{sec:results}
In this section, we present the results of our network when trained on the various datasets.  We implement various procedures to attempt to improve the performance of the network, such as changing the color scaling and optimizing the network architecture hyperparameters for each training dataset.  We also investigate the characteristics of objects commonly classified correctly or incorrectly.

To evaluate each model, we use the receiver operating characteristic (ROC), which calculates the true positive rate (TPR) and false positive rate (FPR) as a function of the threshold score at which the model classifies an object as a lens or non-lens.  From the ROC curve, we calculate the standard metric of the area under the ROC (AUROC) to evaluate each model, as a higher AUROC value corresponds to a better network performance.

\subsection{Network architecture for fiducial dataset} \label{subsec:results_fid}
\begin{figure}
    \centering
	\includegraphics[width=0.5\textwidth]{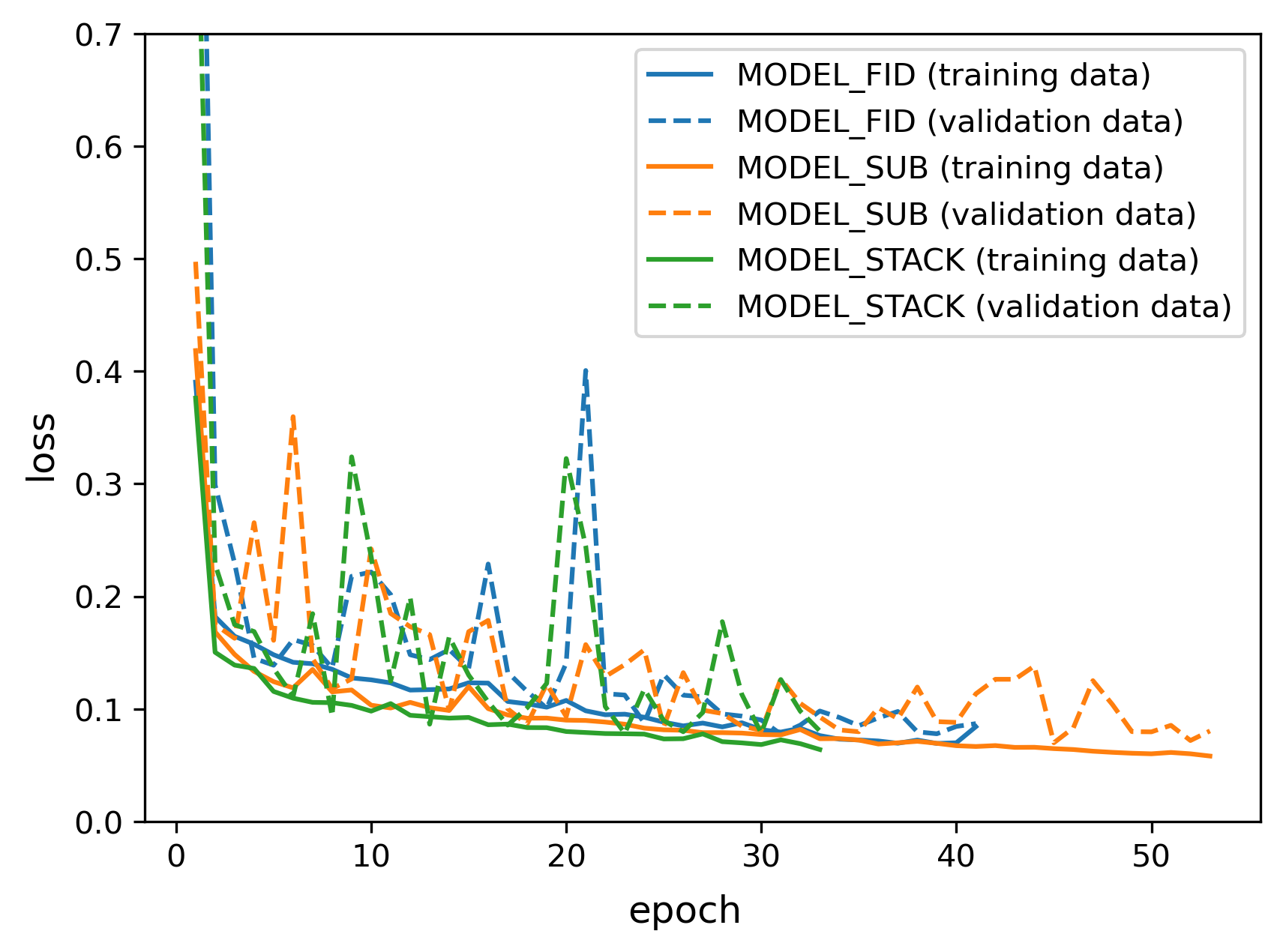}
    \caption{Loss curve of models optimized for the {\sc fid} dataset for the training (solid lines) and validation (dashed lines) datasets.  Shown are the curves for {\sc model\_fid} (blue), {\sc model\_sub} (orange), and {\sc model\_stack} (green).}
    \label{fig:loss_curves_model_fid_color_fid}
\end{figure}

\begin{figure}
    \centering
	\includegraphics[width=0.5\textwidth]{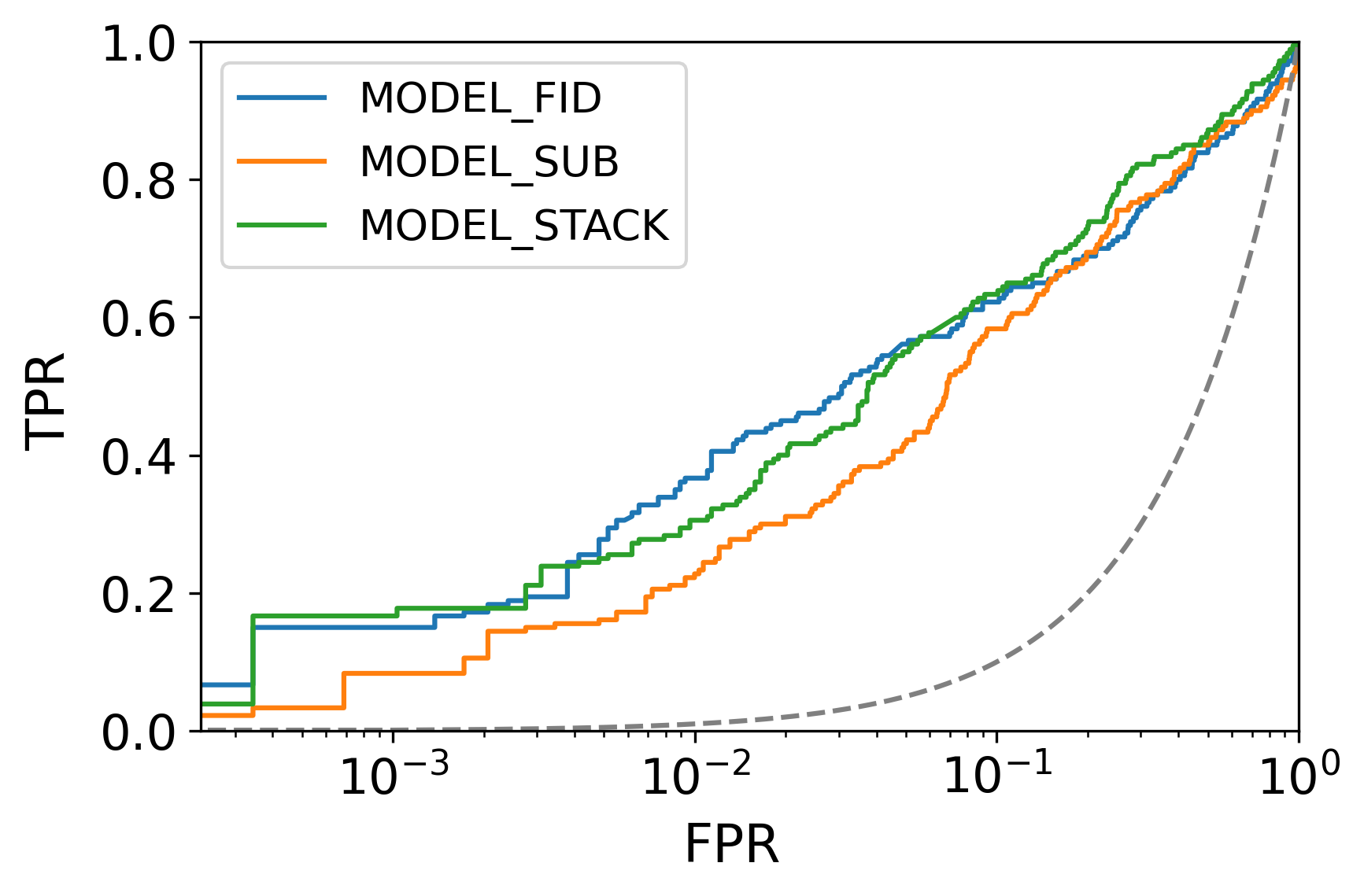}
    \caption{ROC curve of models optimized for the {\sc fid} dataset.  Shown are the curves for {\sc model\_fid} (blue), {\sc model\_sub} (orange), and {\sc model\_stack} (green).  The grey dashed line represents the line where $\mathrm{TPR}=\mathrm{FPR}$.}
    \label{fig:roc_curves_model_fid_color_fid}
\end{figure}

We compare the AUROC of the network optimized (using the Keras Tuner) on the fiducial dataset when trained on each of the three datasets. The AUROC values and the number of epochs of training for each model, denoted ``{\sc model\_fid}", ``{\sc model\_sub}", and ``{\sc model\_stack}", are given in Table~\ref{table: AUROC values of the fiducial network}. The loss curve of the networks optimized on the fiducial dataset are shown in Figure~\ref{fig:loss_curves_model_fid_color_fid}. 
The corresponding ROC curves are shown in Figure~\ref{fig:roc_curves_model_fid_color_fid}.  Based on the AUROC values in Table \ref{table: AUROC values of the fiducial network}, {\sc model\_stack} performs the best, while the values of {\sc model\_fid} and {\sc model\_sub} are slightly lower.  We note that the AUROC value for {\sc model\_fid} (0.808) is an updated result for the ``L1+L2-N1" test sample in Table 2 of \citet{more+2024}, where an earlier version of our network was tested on the same sample and attained an AUROC of 0.80.  Other networks in \citet{more+2024} have higher AUROC values (ranging from 0.86 to 0.98), but this is primarily due to differences in the network architectures and training data.  In this work, since we are evaluating the relative improvement due to light subtraction, it is fair to use our {\sc model\_fid} result as a baseline with which to compare our networks using light subtracted images.  In principle, we could run this same study on the other networks in \citet{more+2024}, but we would still be comparing to their respective fiducial networks to quantify the improvement due to light subtraction.

\begin{table}
    \centering
    \caption{AUROC values of the network with architecture optimized for the {\sc fid} dataset using the Keras Tuner.  The columns show the AUROC value when predicting on the test dataset, the median and standard deviation of the AUROC value over 1000 bootstrap resamplings of the test dataset, and the number of epochs used for training.}
    \label{table: AUROC values of the fiducial network}
    \begin{tabular}{l|r|r|r}
    model&AUROC&AUROC&Epochs\\ 
    & & (bootstrap) & \\
    \hline \hline
    {\sc model\_fid} & 0.808&$0.807 \pm 0.022$&41\\
    {\sc model\_sub} & 0.796&$0.791 \pm 0.023$&53\\
    {\sc model\_stack} & 0.834&$0.833 \pm 0.020$&33\\
    \end{tabular}
    \\
\end{table}

In order to evaluate the uncertainty on the AUROC and whether the differences among the models are significant, we perform a bootstrap resampling of the entire test dataset.  We resample (with replacement) the lenses and non-lenses separately, keeping the number of objects of each type the same in each bootstrap trial, then calculate the AUROC over 1000 realizations.  We show the median and standard deviation of the distribution of AUROC values calculated in this way in Table~\ref{table: AUROC values of the fiducial network}.  We use the same realizations of the test sample for each of the networks for this test.  We find that the AUROC of {\sc model\_stack} is better than the AUROC of {\sc model\_fid} in 960 out of 1000 realizations, and the median AUROC values are very similar to the original model, suggesting that the improvement in network performance when trained on {\sc model\_stack} is a real effect and not simply a statistical anomaly.

\subsection{Optimizing the color scaling} \label{subsec:color_opt}
In preparing the fiducial dataset, we scaled the $g$, $r$, and $i$ band images by multiplicative factors of 2.0, 1.2, and 1.0, respectively (Section~\ref{subsec:data_aug}).  These values were determined through experimentation to find the best network performance.  In principle, the galaxy-subtracted images should remove the central galaxy, which is often a red elliptical galaxy, leaving only the lensed features, which are typically blue.  This may change the optimal multiplicative scaling factors for the {\sc sub} dataset, so we re-optimize them here.  We then construct the updated {\sc stack} dataset by combining the {\sc fid} data with the original scaling factors and the {\sc sub} data with the updated scaling factors.  The AUROC values for these updated data are shown in Table~\ref{table:model_opt}.

\begin{table}
    \centering
    \caption{AUROC values of the networks trained on the {\sc sub} and {\sc stack} datasets when optimizing the color scaling factor and network architecture (using the Keras Tuner), as well as when correcting the central residual of the {\sc sub} dataset.}
    \label{table:model_opt}
    \begin{tabular}{l|r}
    Model&AUROC\\ 
    & (common dataset)\\
    \hline
    \hline
    \multicolumn{2}{c}{Optimized Color Scale} \\
    {\sc model\_sub} & 0.765\\
    {\sc model\_stack} & 0.849\\
    \hline
    \multicolumn{2}{c}{Optimized Network Architecture} \\
    {\sc model\_sub\_opt} & 0.712\\
    {\sc model\_stack\_opt} & 0.837\\
    \hline
    \multicolumn{2}{c}{Central Residual Correction} \\
    {\sc model\_sub\_optcorr}  & 0.820\\
    {\sc model\_stack\_optcorr} & 0.841\\
    \end{tabular}
\end{table}

When we compare the results of models with the color scale optimized for the {\sc sub} dataset with the results of models with the color scale optimized for the {\sc fid} dataset, the performance of {\sc model\_sub} decreases slightly, while the performance of {\sc model\_stack} is slightly increased.  The color scale can have an impact, but the effect is small, and the relative AUROC values of the networks trained on the three datasets remains similar.

\subsection{Optimizing the network architecture for each dataset} \label{subsec:arch_opt}
\begin{figure}
    \centering
	\includegraphics[width=0.5\textwidth]{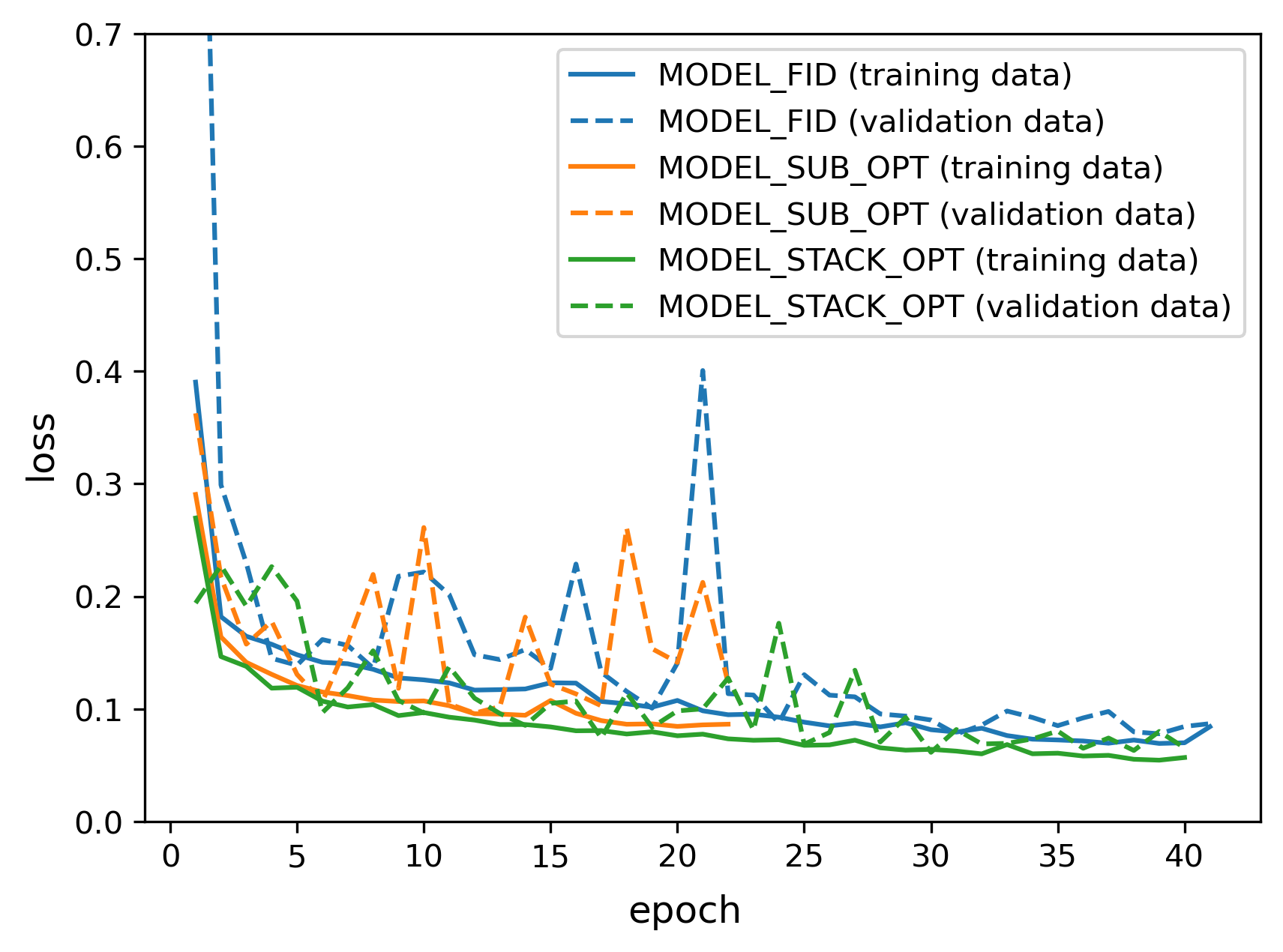}
    \caption{Same as Figure~\ref{fig:loss_curves_model_fid_color_fid}, but for the networks optimized for each respective dataset.}
    \label{fig:loss_curves_model_opt_color_fid}
\end{figure}

\begin{figure}
    \centering
	\includegraphics[width=0.5\textwidth]{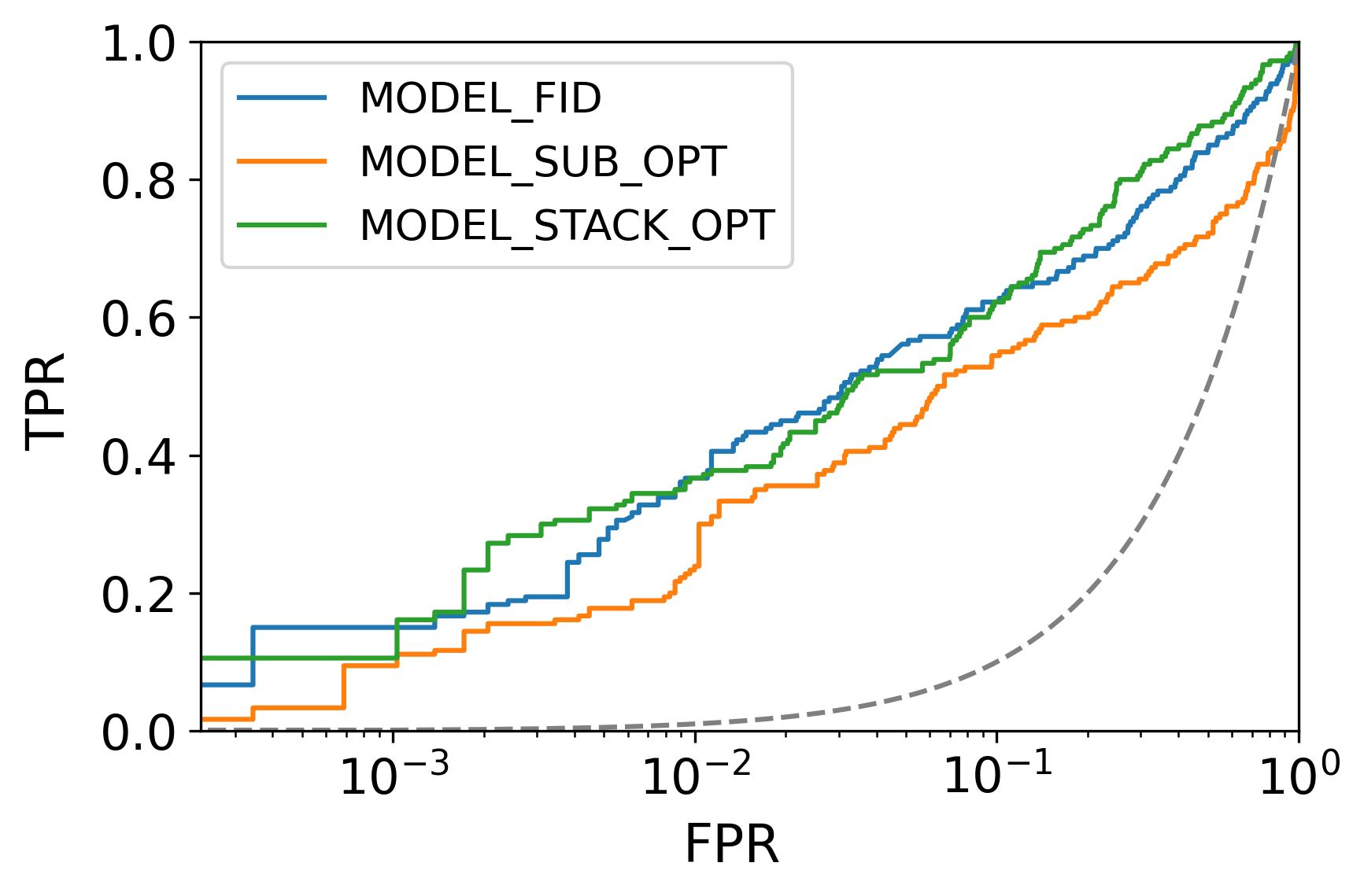}
    \caption{Same as Figure~\ref{fig:roc_curves_model_fid_color_fid}, but for the networks optimized for each respective dataset.}
    \label{fig:roc_curves_opt}
\end{figure}

In addition to optimizing the color scaling, we use the Keras Tuner to optimize the network architecture for each of the the {\sc sub} and {\sc stack} datasets, as described in Section~\ref{sec:cnn}.
The loss curve of each network are shown in Figure~\ref{fig:loss_curves_model_opt_color_fid}, and the ROC curves of each model are shown in Figure~\ref{fig:roc_curves_opt} after this optimization.  The architecture of each network is shown in Table~\ref{table: arch}.  The AUROC values of each model are given in Table \ref{table:model_opt}. {\sc model\_fid} is still the model trained on the dataset which contains the fiducial images.  We denote the models optimized and trained on the {\sc sub} and {\sc stack} datasets as ``{\sc model\_sub\_opt}" and ``{\sc model\_stack\_opt}", respectively, to distinguish them from {\sc model\_sub} and {\sc model\_stack}, which kept the same architecture and color scaling as {\sc model\_fid}.

\renewcommand*\arraystretch{0.85}
\begin{sidewaystable*}
    \centering
    \caption{Architecture of models optimized for each of the training datasets.}
    \label{table: arch}
    \begin{tabular}{|l|l|l|l|l|}
    \hline
    {\sc model\_fid}&{\sc model\_sub\_opt}&{\sc model\_stack\_opt}&{\sc model\_sub\_optcorr}&{\sc model\_stack\_optcorr}\\ 
    Total Parameters = 3,173,825 & Total Parameters = 870,433 & Total Parameters = 1,895,041 & Total Parameters = 1,386,017 & Total Parameters = 1,926,785 \\
    Type(Size, filters, Activ) & Type(Size, filters, Activ) & Type(Size, filters, Activ) & Type(Size, filters, Activ) & Type(Size, filters, Activ) \\
    \hline \hline
    Conv($7\times$7, 64, ReLU) & Conv($7\times$7, 32, ReLU) & Conv($7\times$7, 128, ReLU) &
    Conv($7\times$7, 32, ReLU) &
    Conv($7\times$7, 32, ReLU) \\
    Conv($3\times$3, 64, ReLU) & Conv($3\times$3, 32, ReLU) & Conv($3\times$3, 128, ReLU) &
    Conv($3\times$3, 32, ReLU) &
    Conv($3\times$3, 32, ReLU) \\
    BatchNormalization & BatchNormalization & BatchNormalization & BatchNormalization & BatchNormalization \\
    Conv($3\times$3, 64, ReLU) $\times$ 2 & Conv($3\times$3, 32, ReLU) $\times$ 2 & Conv($3\times$3, 128, ReLU) $\times$ 2 &
    Conv($3\times$3, 32, ReLU) $\times$ 2 &
    Conv($3\times$3, 32, ReLU) $\times$ 2\\
    BatchNormalization & BatchNormalization & BatchNormalization & BatchNormalization & BatchNormalization\\
    MaxPooling(2, NA, NA) & Conv($3\times$3, 32, ReLU) $\times$ 2 & MaxPooling(2, NA, NA) & Conv($3\times$3, 32, ReLU) $\times$ 2 & Conv($3\times$3, 32, ReLU) $\times$ 2 \\
    Conv($3\times$3, 128, ReLU) $\times$ 2 & BatchNormalization & Conv($3\times$3, 128, ReLU) $\times$ 2 & BatchNormalization & BatchNormalization\\
    BatchNormalization & MaxPooling(2, NA, NA) & BatchNormalization & MaxPooling(2, NA, NA) & Conv($3\times$3, 32, ReLU) $\times$ 2 \\
    Conv($3\times$3, 128, ReLU) $\times$ 2 & Conv($3\times$3, 64, ReLU) $\times$ 2 & Conv($3\times$3, 128, ReLU) $\times$ 2 & Conv($3\times$3, 64, ReLU) $\times$ 2 & BatchNormalization \\
    BatchNormalization & BatchNormalization & BatchNormalization & BatchNormalization & MaxPooling(2, NA, NA) \\
    Conv($3\times$3, 128, ReLU) $\times$ 2 & Conv($3\times$3, 64, ReLU) $\times$ 2 & MaxPooling(2, NA, NA) & MaxPooling(2, NA, NA) & Conv($3\times$3, 32, ReLU) $\times$ 2 \\
    BatchNormalization & BatchNormalization & Conv($3\times$3, 128, ReLU) $\times$ 2 &  Conv($3\times$3, 128, ReLU) $\times$ 2 & BatchNormalization \\
    Conv($3\times$3, 128, ReLU) $\times$ 2 & Conv($3\times$3, 64, ReLU) $\times$ 2 & BatchNormalization & BatchNormalization &  MaxPooling(2, NA, NA) \\
    BatchNormalization & BatchNormalization & MaxPooling(2, NA, NA) & MaxPooling(2, NA, NA) & Conv($3\times$3, 128, ReLU) $\times$ 2 \\
    MaxPooling(2, NA, NA) & MaxPooling(2, NA, NA) & Flatten(NA, NA, NA) & Flatten(NA, NA, NA) & BatchNormalization \\
    Conv($3\times$3, 128, ReLU) $\times$ 2 & Conv($3\times$3, 64, ReLU) $\times$ 2 & Dense(64, NA, ReLU) & Dense(128, NA, ReLU) &  Conv($3\times$3, 128, ReLU) $\times$ 2 \\
    BatchNormalization & BatchNormalization & Dropout($\rm{rate}=0.4$) & Dropout($\rm{rate}=0.4$) & BatchNormalization \\
    Conv($3\times$3, 128, ReLU) $\times$ 2 & MaxPooling(2, NA, NA) & Dense(32, NA, ReLU) & Dense(64, NA, ReLU) & Conv($3\times$3, 128, ReLU) $\times$ 2 \\
    BatchNormalization & Flatten(NA, NA, NA) & Dense(1, NA, sigmoid) & Dense(1, NA, sigmoid) & BatchNormalization \\
    Conv($3\times$3, 128, ReLU) $\times$ 2 & Dense(128, NA, ReLU) & & & MaxPooling(2, NA, NA) \\
    BatchNormalization & Dropout($\rm{rate}=0.4$) & & & Flatten(NA, NA, NA) \\
    MaxPooling(2, NA, NA) & Dense(128, NA, ReLU) & & & Dense(128, NA, ReLU) \\
    Flatten(NA, NA, NA) & Dense(1, NA, sigmoid) & & & Dense(64, NA, ReLU) \\
    Dense(128, NA, ReLU) & & & & Dense(1, NA, sigmoid) \\
    Dropout($\rm{rate}=0.4$) & & & & \\
    Dense(64, NA, ReLU) & & & & \\
    Dense(1, NA, sigmoid) & & & & \\
    \hline
    \end{tabular}
\end{sidewaystable*}
\renewcommand*\arraystretch{1.0}

Based on their respective AUROC values, {\sc model\_stack\_opt} performs comparable to {\sc model\_stack}.  However, {\sc model\_sub\_opt} performs worse than {\sc model\_sub}, which was optimized for the {\sc fid} data.
The earlier result of {\sc model\_stack} performing better than {\sc model\_fid} appears to hold with regard to {\sc model\_stack\_opt} as well.

Comparing our results with the results of \citet{canameras+2023}, who looked at difference images, the results are qualitatively similar.  Their CNN, when trained on the stacked fiducial images with difference images, performed better than the network trained on the fiducial dataset alone.  However, the network trained only on the difference images performed worse than the network trained on the fiducial dataset.  Our results support their conclusion, although a quantitative comparison is difficult due to differences in the network architecture and datasets.  \citet{canameras+2023} also tried applying their difference images to a ResNet architecture and found no improvement compared to the fiducial data, but we cannot directly compare our results in this case.

\subsection{Characteristics of objects correctly and incorrectly classified} \label{subsec:objects}

\begin{figure*}
    \centering
	\includegraphics[height=0.9\textwidth]{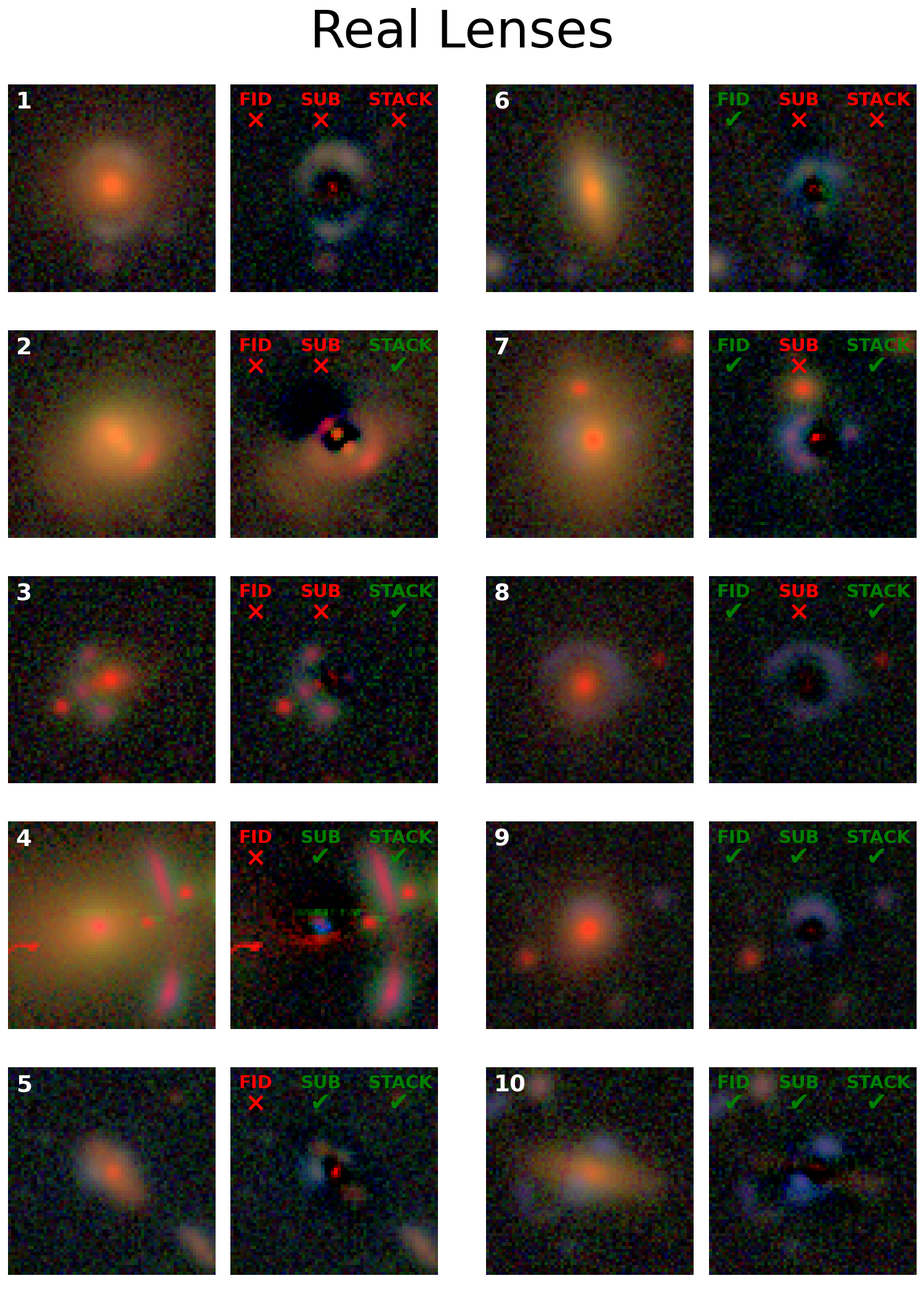}
    \caption{Example images of real lenses in the test sample that are classified correctly or incorrectly by {\sc model\_fid} and {\sc model\_sub\_opt} and {\sc model\_stack\_opt}, as indicated by the green check mark (correct) or red cross (incorrect).  The objects are chosen to be representative of a range of lens configurations.}
    \label{fig:compare_l1}
\end{figure*}

We compare the classifications of real lenses by {\sc model\_fid} with that of {\sc model\_sub\_opt} and {\sc model\_stack\_opt} in Figure~\ref{fig:compare_l1}.  The example objects are chosen to be representative of the different types of objects generally classified correctly or incorrectly by the respective networks.  We choose thresholds for each network to maximize $\rm{TPR}-\rm{FPR}$.   When we compare the objects indicated by panels 4 and 5 with those indicated by panels 9 and 10, the lens configurations are similar. However, {\sc model\_fid} incorrectly classifies objects 4 and 5 as non-lenses. The light from the lensed images in panels 4 and 5 is fainter than that of panels 9 and 10, which might be the cause of the misclassifications. On the contrary, the light from the lensed images in the subtracted images of these objects is highlighted when the lensing galaxy light is removed, so {\sc model\_sub\_opt} and {\sc model\_stack\_opt} classify these images correctly.
In addition, both of {\sc model\_fid} and {\sc model\_sub\_opt} classify the object in panel 3 incorrectly, but {\sc model\_stack\_opt} classifies this image correctly. Therefore, the combination of {\sc fid} data and {\sc sub} data can improve the classification of certain images. 

On the other hand, when we look into the images that {\sc model\_fid} classifies correctly and {\sc model\_sub\_opt} classifies incorrectly, we see that there are often red residual features of the light subtraction in the center of these images (i.e., panel 7).  We investigate this further in Section~\ref{subsec:sub_resid_results}.  On rare occasions, {\sc model\_stack\_opt} classifies an object incorrectly but either {\sc model\_fid} or {\sc model\_sub} correctly identifies it as a lens.  Panels 2 and 6 show such examples, and both appear to be an unusual system with a red source, which we expect to be rare.

\begin{figure*}
    \centering
	\includegraphics[height=0.9\textwidth]{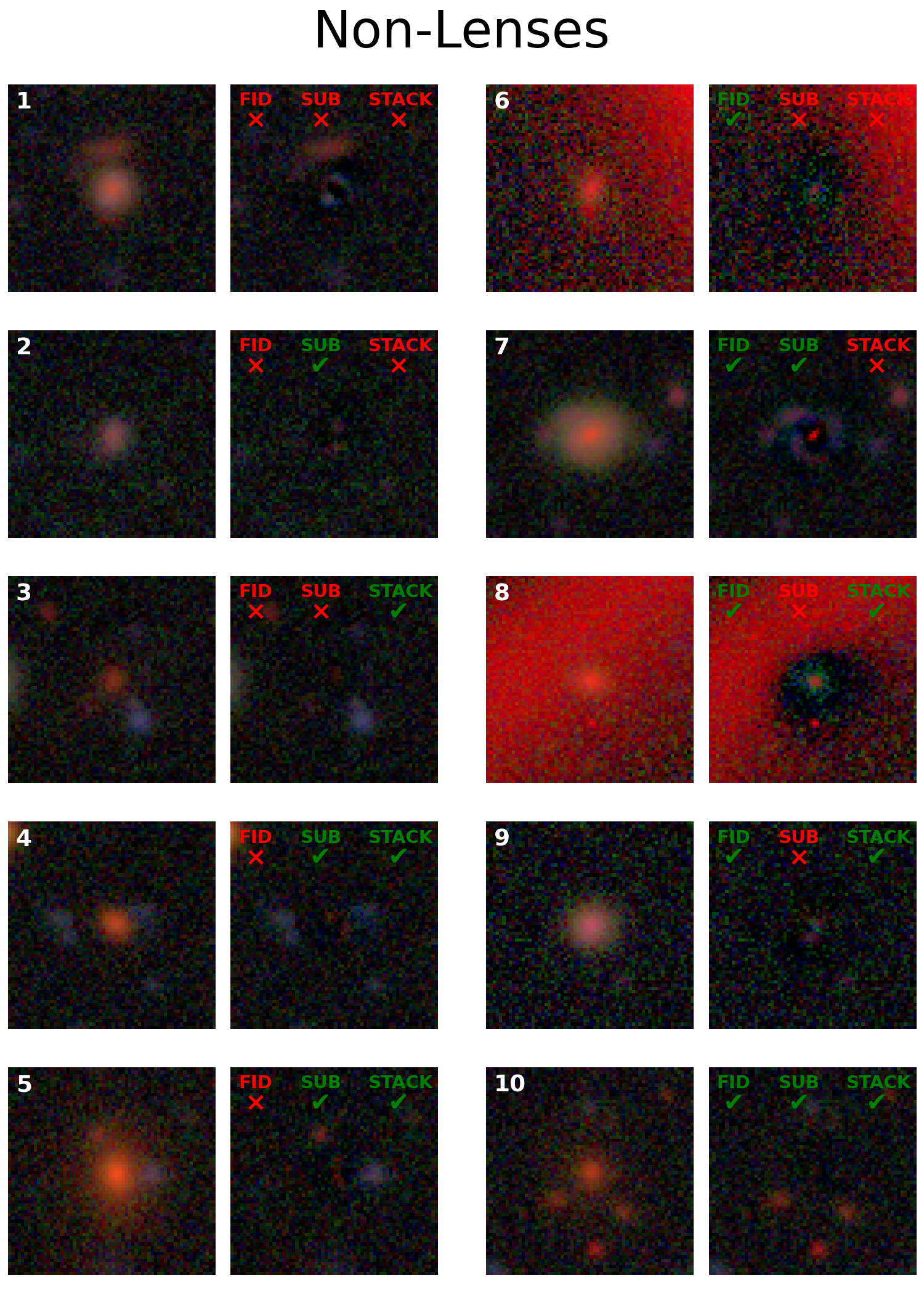}
    \caption{Same as Figure~\ref{fig:compare_l1}, but for real non-lens objects in the test sample.}
    \label{fig:compare_n1}
\end{figure*}

In Figure~\ref{fig:compare_n1}, we compare classifications of non-lenses in the test sample.  Objects such as the one shown in panel 1 are incorrectly classified as a lens by all networks because there can be blue features that can mimic a lens configuration.  This is unlikely to be a real lens given the thinness of the blue features in the right panel relative to the PSF size, indicating that it may be an artifact of the light subtraction.  There are many structures similar to the multiple images of lens systems in the images that {\sc model\_fid} incorrectly interprets as lensed images but that {\sc model\_sub\_opt} and {\sc model\_stack\_opt} identifies correctly, such as galaxies along the line of sight that are not lensed.  These structures are highlighted in the subtracted image, and it is easier to identify such structures as non-lensing features due to their asymmetry, so using the subtracted data leads to a correct classification.  Some objects have photometric artifacts, such as the objects in panels 6 and 8, so {\sc model\_sub\_opt} classifies these images incorrectly.  {\sc model\_stack\_opt} is able to correctly classify the object in panel 8 on which {\sc model\_sub\_opt} made a mistake, providing an example of how combining the two datasets can overcome certain misclassifications.

\subsection{Removing central residual from subtracted data} \label{subsec:sub_resid_results}
\begin{figure}
    \centering
	\includegraphics[width=0.5\textwidth]{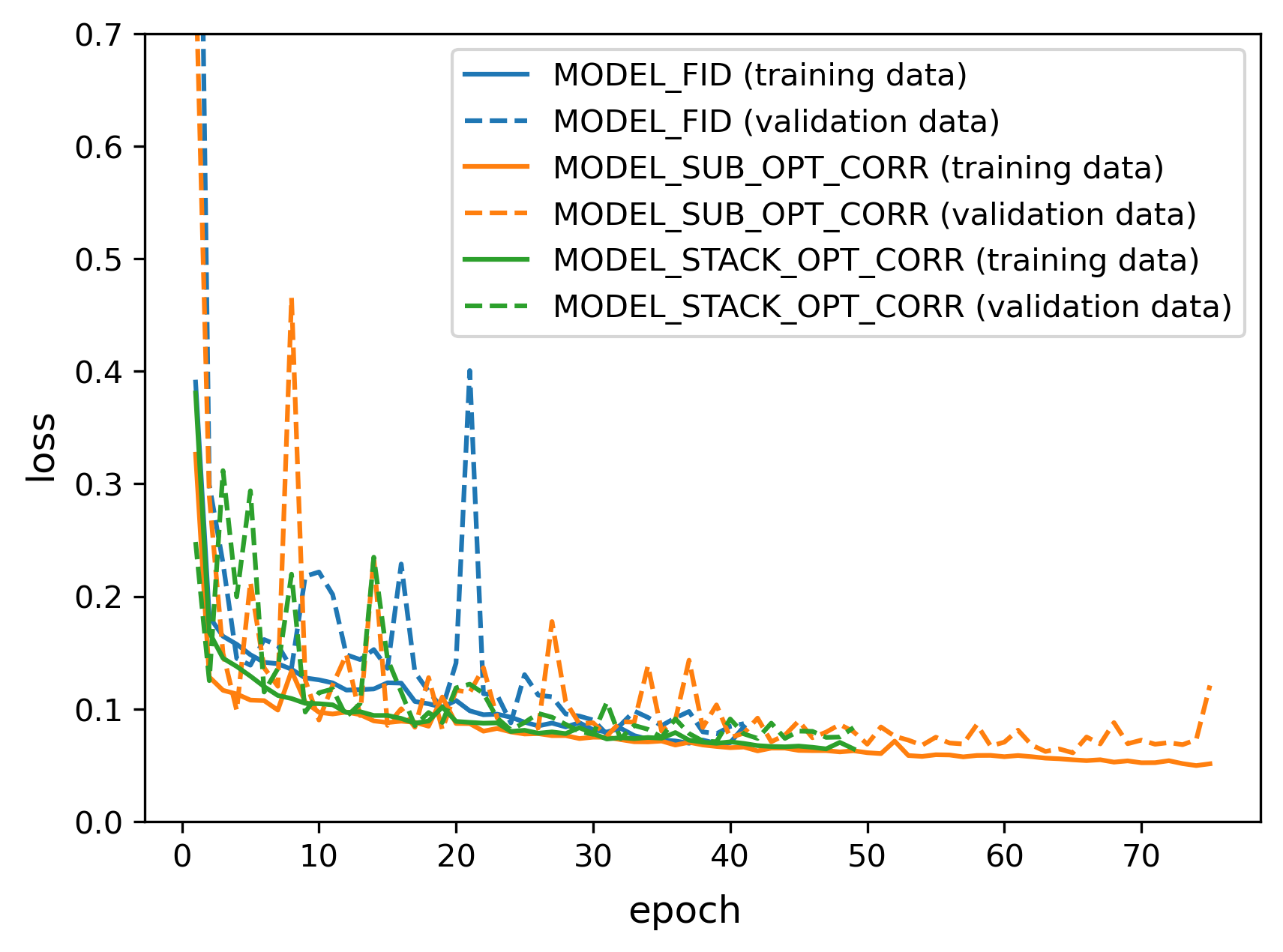}
    \caption{The loss curves of {\sc model\_sub\_opt} (blue solid line) and {\sc model\_stack\_opt} (orange solid line) compared to the networks with the central residual correction applied: {\sc model\_sub\_optcorr} (blue dashed line) and {\sc model\_stack\_optcorr} (orange dashed line).}
    \label{fig:loss_curve_residual_correction}
\end{figure}

We further investigate whether the red residual features of the light subtraction in the center of the images (Section~\ref{subsec:objects}) could be causing objects to be incorrectly classified in the {\sc sub} and/or {\sc stack} datasets (e.g., Figure~\ref{fig:compare_l1}, panel 7; Figure~\ref{fig:compare_n1}, panel 7).  We attempt to correct this central residual feature by replacing the affected pixels in the subtracted image with noise as follows.  We identify a pixel as being associated with a galaxy if the fluxes of the pixel in a particular band in the fiducial image is larger than 0.3.  We then calculate the median $\mathrm{med}(x)$ and the normalized median absolute deviation $\sigma(x)$ from all pixels of the image which are not in a galaxy, as a way of estimating the background noise level.  We then replace the fluxes of pixels in each band in the subtracted images with Gaussian noise with scale $\mathrm{med}(x)$ and standard deviation $\sigma(x)$ if that pixel satisfies at least one of the following criteria:
\begin{enumerate}
\item $i>0.1$ {\sc and} $i > g \times 2$ {\sc and} $i > r \times 2$
\item $r>0.1$ {\sc and} $r > g \times 2$ {\sc and} $r > i \times 2$,
\end{enumerate}
where $g$, $r$, and $i$ are the fluxes of each pixel in the respective bands.  These criteria were determined through testing and were found to most successfully reduce the central residual feature.  We apply this procedure to the pixels in a 8 pixel $\times$ 8 pixel square in the center of each subtracted image.  We then retrain and re-optimize the color scale and network architecture using these updated datasets with corrected central residuals.  We call these networks {\sc model\_sub\_optcorr} and {\sc model\_stack\_optcorr}.  The loss curves are shown in Figure \ref{fig:loss_curve_residual_correction}, and do not show any indication that the network is overfitting to the training data.  The AUROC values from these tests are given in Table~\ref{table:model_opt}.

\begin{figure}
    \centering
	\includegraphics[width=0.5\textwidth]{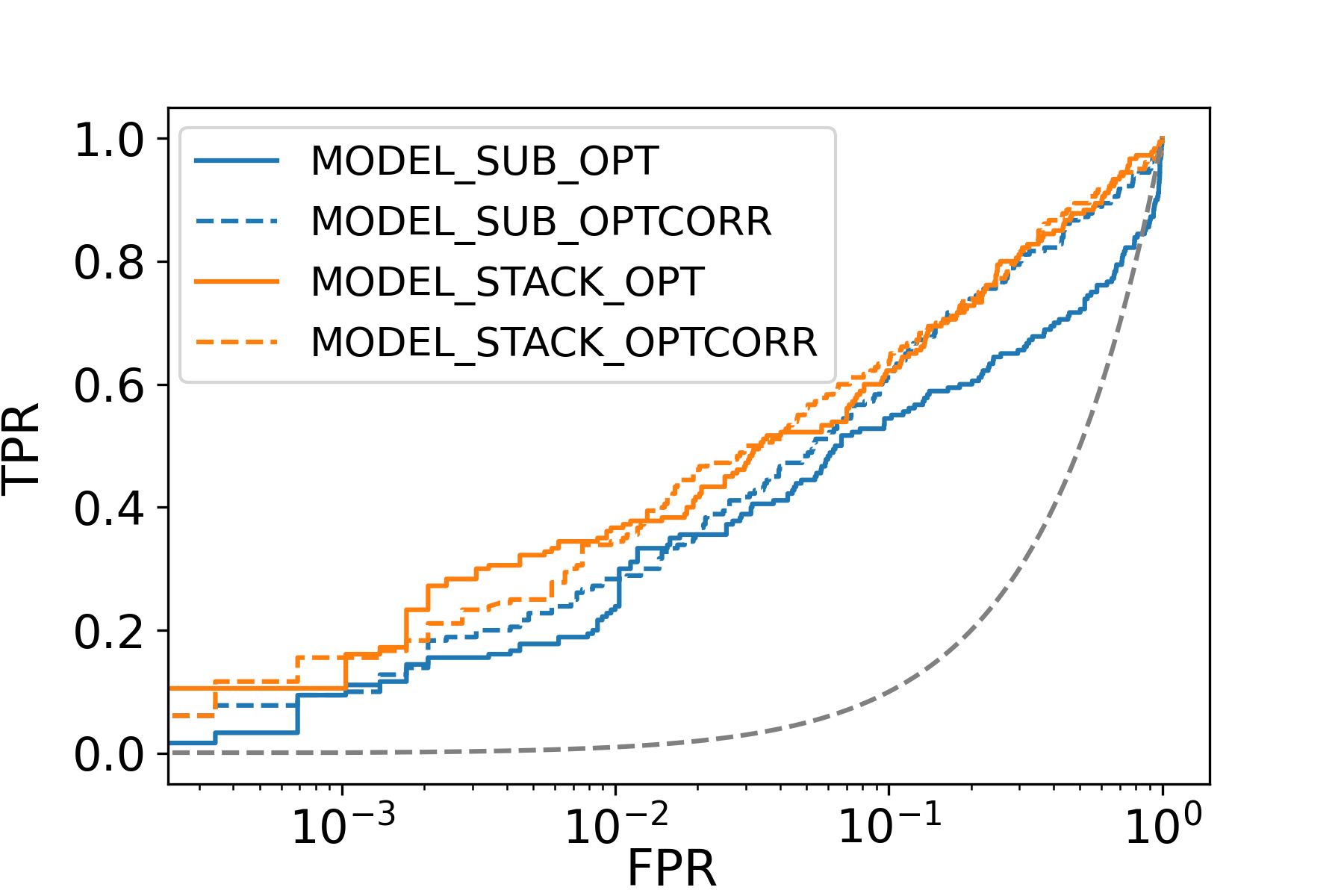}
    \caption{The ROC curves of {\sc model\_sub\_opt} (blue solid line) and {\sc model\_stack\_opt} (orange solid line) compared to the networks with the central residual correction applied: {\sc model\_sub\_optcorr} (blue dashed line) and {\sc model\_stack\_optcorr} (orange dashed line).}
    \label{fig:roc_curve_residual_correction}
\end{figure}

When we compare the results of models with this central residual correction with the results of models without the correction, the performances of {\sc model\_sub\_optcorr} is substantially improved compared to {\sc model\_sub\_opt} (Figure~\ref{fig:roc_curve_residual_correction}).  Therefore, the central residual appears to have been a significant factor in the lower performance of {\sc model\_sub\_opt}.  However, the performance of {\sc model\_stack\_optcorr} remains similar to that of {\sc model\_stack\_opt}, indicating that the combination of the fiducial imaging data with the subtracted data is still a larger factor in the improved classification.  More sophisticated light subtraction methods may provide a more clear improvement.

\section{Summary and future work} \label{sec:summary}
In this work, we develop a convolutional neural network to find galaxy-scale strong lenses that is trained on data that combines the original images with images that have the central galaxy light profile subtracted.  Given that typical image separations are comparable to the atmospheric seeing, subtracting the lens galaxy light helps to highlight the light from the images of the source galaxy.  This approach and its benefits is not specific to our network, and may also improve the performance of previous and future CNN-based lens searches, although fast light subtraction algorithms will be needed for increasingly large datasets.

We make use of the HSC SSP $gri$-band data, which not only provides the best image resolution and depth among current wide ground-based surveys, but is also a precursor to the upcoming Rubin/LSST.  We make three datasets: the fiducial multiband imaging data, the light-subtracted data, and the stack of both the fiducial and light-subtracted data.  We compare the results of training and testing the three datasets, keeping the architectures of the networks same and also optimizing the architectures of the networks for each dataset.  Our main conclusions are as follows:
\begin{enumerate}
    \item Training the CNN and applying it to the light-subtracted images ({\sc sub} dataset) themselves does not necessarily improve the overall performance of classification compared to the fiducial images ({\sc fid} dataset), although it can correctly classify certain objects that were incorrectly classified with the fiducial data (and vice versa).
    \item Combining the fiducial imaging data with the subtracted data ({\sc stack} dataset) does seem to improve network performance, as the AUROC values of {\sc model\_stack\_opt} is better than the AUROC values of {\sc model\_fid} and {\sc model\_sub\_opt}.  This indicates that combining information in this way can strike a balance between highlighting the characteristics of the light from source galaxy and using the features of the original image.  This is consistent with the results of \citet{canameras+2023}, who used difference images as a proxy for galaxy light subtraction.  This approach can be applied to improve the performance of classification for future lens searches with CNNs.
    \item Correcting the residual features of the light subtraction from the {\sc sub} dataset can improve the performance of classification, indicating that improving the light subtraction scheme can generally improve performance.  It is unclear whether this is a larger factor than stacking the fiducial imaging data and subtracted data though, as there was minimal performance gain when applied to {\sc model\_stack\_opt}.  Future exploration of alternate methods of separating the central (lens) galaxy light from the lensed background source may further improve classification.
\end{enumerate}

In the future, we can further investigate the merits of our work by applying it to different sets of training data (e.g., \cite{canameras+2021}; \cite{shu+2022}) to see if our conclusions can be generalized, or if the choice of a specific training set has an effect.  Using ensemble classifiers (e.g., \cite{holloway+2024}) can improve the accuracy of classification, making the process more time-efficient.

We can also investigate more sophisticated galaxy light subtraction algorithms (e.g., \cite{joseph+2016,savary+2022}; Rojas et al. in preparation) that may do a better job at modeling and removing the central galaxy light.  The current computational efficiency of the light subtraction step was sufficient for our study, but is not feasible for future datasets (e.g., LSST) that may have on the order of $\sim10^{8}$ objects to classify.  A more detailed study about the tradeoffs between light profile fitting/subtraction accuracy and runtime, and how they impact network performance, may lead to comparable improvements while reducing the computational resources needed.  Newer algorithms that can make use of GPU parallelization for galaxy light profile fitting (e.g., \cite{chen+2024}) are also promising avenues to explore in the future.

This methodology can be applied to the same sample classified by \citet{jaelani+2023} to discover galaxy-scale lenses from the entire HSC SSP PDR2 dataset.  Such an analysis will allow us to understand the true improvement brought about by our light subtraction methodology.  If proven to be more efficient, this will further motivate the implementation of such methodologies in upcoming surveys such as LSST.

\begin{ack}
We thank Alessandro Sonnenfeld for useful discussions regarding the {\sc YattaLens} software.
We thank Raoul Ca\~{n}ameras, Karina Rojas, and Masayuki Tanaka for useful discussions and input.
This work is supported by JSPS KAKENHI Grant Numbers JP20K14511, JP24K07089.
A.T.J. is supported by the Riset Unggulan ITB 2024.
The Hyper Suprime-Cam (HSC) collaboration includes the astronomical communities of Japan and Taiwan, and Princeton University. The HSC instrumentation and software were developed by the National Astronomical Observatory of Japan (NAOJ), the Kavli Institute for the Physics and Mathematics of the Universe (Kavli IPMU), the University of Tokyo, the High Energy Accelerator Research Organization (KEK), the Academia Sinica Institute for Astronomy and Astrophysics in Taiwan (ASIAA), and Princeton University. Funding was contributed by the FIRST program from the Japanese Cabinet Office, the Ministry of Education, Culture, Sports, Science and Technology (MEXT), the Japan Society for the Promotion of Science (JSPS), Japan Science and Technology Agency (JST), the Toray Science Foundation, NAOJ, Kavli IPMU, KEK, ASIAA, and Princeton University.
This paper is based on data collected at the Subaru Telescope and retrieved from the HSC data archive system, which is operated by Subaru Telescope and Astronomy Data Center (ADC) at NAOJ. Data analysis was in part carried out with the cooperation of Center for Computational Astrophysics (CfCA) at NAOJ. We are honored and grateful for the opportunity of observing the Universe from Maunakea, which has the cultural, historical and natural significance in Hawaii.
This research made use of \textsc{NumPy} \citep{oliphant2015,harris+2020} and \textsc{SciPy} \citep{jones+2001,virtanen+2020}. This research made use of \textsc{matplotlib}, a 2D graphics package used for \textsc{Python} \citep{hunter2007}.
\bibliographystyle{myaasjournal}
\bibliography{cnn_lightsub}
\end{ack}

\appendix
\section*{Earlier version of network} \label{app:prev}
As the networks presented in this paper were being developed, a preliminary version of the CNN trained on the {\sc fid} dataset was included in \citet{holloway+2024} and \citet{more+2024} as part of their work comparing the performance of various lens-finding networks.  This preliminary network was built before implementing the Keras Tuner, but is still a classical CNN with a similar structure.  For completeness, we present the architecture of this earlier version in Table~\ref{tab:old_model}.  

\renewcommand*\arraystretch{0.85}
\begin{table}[htbp]
    \centering
    \caption{Diagram showing the architecture of an earlier version of our network that was presented in \citet{holloway+2024} and \citet{more+2024}.}
    \label{tab:old_model}
    \begin{tabular}{|l|}
    \hline
    model\_fid\\
    Total Parameters = 1,792,673\\
    Type(Size, filters, Activ)\\
    \hline \hline
    Conv($7\times$7, 32, ReLU)\\
    Conv($3\times$3, 32, ReLU)\\
    BatchNormalization\\
    MaxPooling(2, NA, NA)\\
    Conv($3\times$3, 64, ReLU) $\times$ 2\\
    BatchNormalization\\
    Conv($3\times$3, 64, ReLU) $\times$ 2 \\
    BatchNormalization\\
    Conv($3\times$3, 64, ReLU) $\times$ 2 \\
    BatchNormalization\\
    MaxPooling(2, NA, NA)\\
    Conv($3\times$3, 128, ReLU) $\times$ 2 \\
    BatchNormalization \\
    Conv($3\times$3, 128, ReLU) $\times$ 2 \\
    BatchNormalization \\
    MaxPooling(2, NA, NA)\\
    Flatten(NA, NA, NA)\\
    Dense(128, NA, ReLU)\\
    Dropout($\rm{rate}=0.4$)\\
    Dense(64, NA, ReLU)\\
    Dropout($\rm{rate}=0.4$)\\
    Dense(1, NA, sigmoid)\\
    \hline
    \end{tabular}
\end{table}
\renewcommand*\arraystretch{1.0}

\end{document}